# Evolution Process of Wurtzite ZnO Films on Cubic MgO (001) Substrates: a Structural, Optical and Electronic Investigation of the Misfit Structures

Hua Zhou[1], Hui-Qiong Wang[1]*, Yaping Li[1], Junyong Kang[1], Jin-Cheng Zheng[1],

Zheng Jiang[2], Yuying Huang[2], Lijun Wu[3], Lihua Zhang[4], Kim Kisslinger[4], Yimei Zhu[3]

[1]*Key Laboratory of Semiconductors and Applications of Fujian Province, Department of Physics, Xiamen University, Xiamen 361005, P. R. China*

[2]*Shanghai Synchrotron Radiation Facility, Shanghai Institute of Applied Physics, Chinese Academy of Sciences, Shanghai 201800, P. R. China*

[3]*Condensed Matter Physics and Materials Science Department, Brookhaven National Laboratory, Upton, New York 11973, USA*

[4]*Center for Functional Nanomaterials, Brookhaven National Laboratory, Upton, New York 11973, USA*

*Corresponding Author. Email: hqwang@xmu.edu.cn

## ABSTRACT

The interface between hexagonal ZnO films and cubic MgO (001) substrates, fabricated through molecular beam epitaxy, are thoroughly investigated. X-ray diffraction and (scanning) transmission electron microscopy reveal that, at the substrate temperature above 200℃, the growth follows the single [0001] direction; while at the substrate below 150℃, the growth is initially along [0001] and then mainly changes to [0-332] variants beyond the thickness of about 10 nm. Interestingly, a double-domain feature with a rotational angle of 30° appears for the growth along [0001] regardless of the growth temperature, experimentally demonstrated the theoretical predictions for occurrence of double rotational domains in such a heteroepitaxy [Grundmann et al, Phys. Rev. Lett. **105**, 146102 (2010)]. It is also found that, the optical transmissivity of the ZnO film is greatly influenced by the mutation of growth directions, stimulated by the bond-length modulations, as further determined by X-ray absorption Spectra (XAS) at Zn K edge. The XAS results also show the evolution of $4p_{xy}$ and $4p_z$ states in the conduction band as the growth temperature increases. The results obtained from this work can hopefully promote the applications of ZnO in advanced optoelectronics for which its integration with other materials of different phases is desirable.

## I. INTRODUCTION

ZnO, crystallizing in the wurtzite structure, is a typical semiconductor with direct and wide band gap as well as large free-exciton binding energy, whereas MgO is an insulator exhibiting the simple rock-salt cubic structure. The integration of ZnO and MgO is motivated by several regards. On one hand, band-gap engineering on wurtzite ZnO can be realized through alloying with cubic MgO; on the other hand, the growth of wurtzite ZnO on the MgO substrates can provide some guidelines for the integration of wurtzite ZnO films with more complicated perovskite cubic substrates, which is desirable for the integration of semiconductor-based "active" devices and perovskite-based "passive" devices. There are already several reports [1-5] about ZnO thin films grown on MgO (001) substrates. Interestingly, ZnO films can be grown along different crystalline directions on the same MgO (001) substrates under different growth conditions or using different growth methods. [6-22] For instance, both oxygen partial pressure[14-18] and substrate temperature [19-22] are found to influence the growth orientations. Nevertheless, the mechanism of the growth orientation is not yet conclusive. In this work, the atomic interfacial structure between the polar ZnO plane and the MgO (001) substrate as well as the evolution of ZnO thin films from polycrystallization with multiple growth directions to single crystallization along the polar growth direction are investigated using high resolution transmission electron microscopy (HRTEM) and scanning transmission electron microscopy (STEM), which has been seldom reported. In addition, the evolution of electronic structure of the ZnO thin films is monitored through X-ray absorption fine structure (XAFS) spectra, which shows no obvious evidence of localization of the conduction band exhibited in amorphous ZnO films. The band-gap structure changes of the ZnO films are also traced and discussed.

## II. EXPERIMENTAL AND COMPUTATIONAL DETAILS

The ZnO films were grown on the single-crystal MgO (001) substrates in an OMICRON MBE system. Before inserted into the ultra-high vacuum chamber, the substrates were cleaned in the ultrasonic baths of acetone and ethanol subsequently. The substrate surfaces were then thermally treated at the temperature of 350°C for an hour, in the oxygen plasma with a partial pressure of $5\times10^{-5}$ mbar and a power of 250 W. The annealed MgO (001) surface showed streaky-like (1×1) RHEED patterns (not shown), similar to the cases in which the surface was either treated at the temperature of 420°C for 60 mins or at the temperatures of 250°C and 350°C each for 60 mins.[19, 23] ZnO films were deposited at a range of substrate temperature with an oxygen partial pressure of $1\times10^{-5}$ mbar and a plasma power of 250 W, while the evaporation rate of Zn (with a purity of 99.9999%) was maintained at 3 nm/min. The texture orientation of the grown ZnO film was characterized using the Panalytical X'pert PRO X-ray diffraction facility. Peak positions for the substrates were used as references to calibrate the peak positions for each sample. The interface structure of ZnO (0001) / MgO (001) and the evolution of the ZnO thin films structure were examined using the JEOL2100F and JEM-ARM200CF equipments for HRTEM and STEM, for which the focused ion beam (FIB) in situ lift-out technique was used to prepare the TEM thin film samples. The electronic structures of the ZnO (0001) thin films were probed through X-ray Absorption Spectra (including XANES and EXAFS) collected on the BL14W1 beamline at Shanghai Synchrotron Radiation Facility (SSRF), Shanghai Institute of Applied Physics (SINAP), China. The incident x-ray energy was selected through Si(111) double-crystal monochromator. The optical property of thin films was examined by the transmission spectra.

The band structure, density of states (DOS) and theoretical XANES spectra of ZnO were calculated using first principles full-potential calculations based on density functional theory (DFT), [24] as implemented in WIEN2k package.[25] The Perdew-Burke-Ernzerh of generalized gradient approximation (PBE-GGA)[26] was used for exchange-correlation function. The Extended X-ray Absorption Fine Structure (EXAFS) analysis and fitting were performed using the IFEFFIT package.[27-29]

## III. RESULTS AND DISCUSSIONS
### A. Structural evolution of (0001) ZnO / MgO (001)

Figure 1 shows the X-ray Diffraction (XRD) result of a series of 60 min-growth ZnO films on the MgO (001) surface, corresponding to the substrate temperatures of 100 °C, 150 °C, 200 °C, 250 °C and 300 °C, respectively. At the substrate temperature range of 200 °C to 300 °C, the strong intensity of the 0002 peak indicates the main (0001) orientation of the ZnO epilayer as shown in our earlier report, [19] notwithstanding for the substrate temperature below 150 °C, while an additional peak of 10-11 appears, as indicated by the arrow in Fig. 1. Nevertheless, the apparently weaker intensity of the 10-11 peak compared to that of the 0002 peak does not necessarily indicate smaller portion of the tilted growth compared to the growth along [0002] direction. One reason is that, structure factor also plays an important role in determining the peak intensity. Furthermore, the tilted growth direction is actually most likely along about [0-332] instead of [0-111]* (the normal of (0-111) plane) as determined from STEM results (see Fig. 4(b) and the discussion later), whose corresponding azimuth deviates about 3° from the [0-111]*, well consistent with the XRD pole figure of {10-11} (2theta=36.2°, note, 10-11 and 0-111 overlap in XRD) reflections (not shown here). Therefore, the 10-11 spot is off Bragg position in XRD, resulting in its future decrease of intensity as shown in Fig. 1. In order to investigate the effect of substrate temperature on the structure of the ZnO films, the mean grain sizes of the five samples are calculated using the Scherrer equation:[23, 30] $D = 0.94\lambda /(B\cos\theta)$, where $D$, $\lambda$, $B$ and $\theta$ are the mean grain size, the X-ray wave length (1.54 Å), the full width at half maximum (FWHM) of the peak in radians, and the Bragg diffraction angle, respectively. As shown in Table I, the calculated mean grain sizes of the five film samples are approximately 11.1 nm, 18.0 nm, 30.9 nm, 36.2nm, 28.8 nm, respectively, which indicates that the change tendency of the grain sizes is first increasing and then decreasing.

TABLE I. XRD FWHM values of the ZnO films grown at different temperature and the corresponding grain sizes. (T: Temperature)

| Substrate T | 100 °C | 150 °C | 200 °C | 250 °C | 300 °C |
|---|---|---|---|---|---|
| FWHM (°) | 0.7763 | 0.4753 | 0.2805 | 0.2955 | 0.2393 |
| Grain size (nm) | 11.1 | 18.0 | 30.9 | 28.8 | 36.2 |

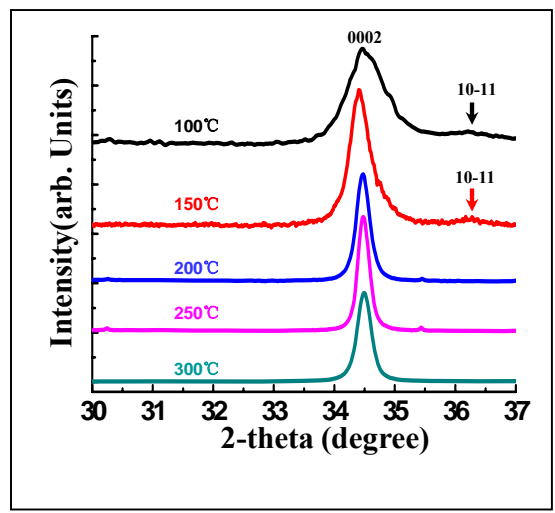

FIG. 1. XRD data for the grown ZnO thin films showing the highly [0001] c-axis oriented texture for all temperatures, with additional 10-11 XRD peaks for films grown at 100 °C and 150 °C.

Figures 2 and 3 show the high resolution transmission electron microscopy (HRTEM) images captured along [110] MgO azimuth from the ZnO films grown at 250℃ and 100℃, respectively. The thickness of the 250℃ sample is about 80 nm, as determined from the low magnification TEM image in Fig. 2(a). The corresponding selected-area electron diffraction (SAED) pattern from the 250℃ sample is shown in Fig. 2(c), which reveals two sets of reciprocal lattices on the zone axis of MgO [110] (the SAED of the substrate is not shown), labeled as white solid and white dashed boxes, corresponding to [2-1-10] domain (named as "m-domain") and [1-100] domain (named as "a-domain"), respectively. The experimental electron diffraction patterns are in agreement with those simulated, as shown in the upper-left inset of Fig. 2(c). The lower-right inset shows the corresponding reflection high energy electron diffraction (RHEED) pattern captured in-situ immediately after growth, which indicates two sets of reciprocal lattices (labeled as $[2a*, 2c*]$ and $[\frac{2}{\sqrt{3}}a*, c*]$, respectively) for the side view of epilayer, corresponding to the basic vectors of the reciprocal lattice for m-plane, and a-plane of ZnO, respectively. The RHEED and SAED results are in excellent agreement. Besides, the same RHEED patterns reappear at each 30° rotation of the sample, which further demonstrates the coexistent domains rotating 30° from each other. In fact, for other samples of ZnO films grown along [0001] direction, the RHEED patterns (not shown here) exhibit the same features, indicating that the interfacial structure of all the samples grown along [0001] azimuth is the same as mentioned above. Furthermore, the SAED pattern on the zone axis of MgO [110] in Fig. 2(c), together with the fact that the aforementioned RHEED patterns (the lower-right inset in Fig. 2(c)) always appear along the [110] direction of MgO, indicates that the registry relationship between the thin film and the substrate based on Figs. 2 and 4 should be:

(0001)ZnO ∥ (001)MgO；(01-10)ZnO ∥ (-110)MgO；[2-1-10]ZnO ∥ [110]MgO and

(0001)ZnO ∥ (001)MgO；(11-20)ZnO ∥ (1-10)MgO；[1-100]ZnO ∥ [110]MgO.

The observed RHEED patterns are similar to those of ZnO (0001) films grown on the MgO (001) substrates with slightly different growth temperature, oxygen partial pressure and plasma power, [19, 31] indicating the robustness of the double-domain (0001) surface structure, for substrate temperature lower than 300 ℃. Hexagonal ZnO (0001) is of six fold symmetry, and cubic MgO (001) is of four fold symmetry. The integration of both leads to two rotational domains in the epilayer. This is due to the mismatch of rotational symmetry at the interface as suggested by the following formula: [32]

$$N_{RD} = \frac{lcm(n,m)}{m},$$

where $N_{RD}$ is the number of rotational domains expected in the grown epilayer, $n$ denotes the $Cn$ rotational symmetry of the substrate crystal with rotation angles $\Phi_i=2\pi/n$, $m$ denotes the $C_m$ rotational symmetry of the epilayer crystal with rotation angles $\Phi_j=2\pi/m$, $lcm(n, m) = k/(i/n+j/m)$ is the least common multiple of $n$ and $m$, where $i$, $j$ and $k$ are integers. In our case, the cubic bulk MgO and hexagonal cubic ZnO have rotational symmetries of $C_4$ $(n=4)$ and $C_6$ $(m=6)$, respectively. And $lcm(4,6)=12$. Therefore, $N_{RD}=12/6=2$, *e.g.* two rotational domains are expected in the hexagonal ZnO epilayer grown on the cubic MgO substrate, which is exactly what our results have shown.

Interestingly, previous theoretical report [33] showed that the boundary energy of the sample with a rotation angle of 30° is larger than that with a 32° rotation. Our observation of a stable rotation angle of 30° could be attributed to the possibility that the interfacial energy between the thin films and the substrate for the rotation angle of 30°, with crystallographically equivalent interfaces in hetero-epitaxy, is lower than that for the rotation of 32°, with enough energy to offset the increment of the boundary energy. Figure 2(d) shows the typical high resolution TEM images from the interface of the ZnO film and the MgO substrate and that of the rotational domains, indicating structural modulation across the interface. Figure 2(e), (f), (g), (h) and (j) are the corresponding amplified images labeled by rectangular areas from the ZnO film. Figures 2(e) and 2(f), corresponding to the regions of the domain m and domain a, respectively, show their atomic structures. The values of structural parameters, *a* and *c*, of the ZnO thin film are measured to be 3.255 Å and 5.243 Å, respectively, which are only about 0.2% and 0.7% larger than those of the ideal bulk ZnO ($a$ = 3.249 Å and $c$ = 5.207 Å), respectively, revealing a nearly fully relaxed thin film. However, the bonding configuration at the interface becomes complicated since (0001) ZnO plane is polar, with Zn and O atoms stacking in alternative layers, while in the MgO (001) plane, there coexist both Mg and O atoms with one dangling bond from each atom. Therefore the interfacial structure will not undergo a simple and direct transition from the MgO (001) surface to the ZnO (0001) polar surface, without any extra dangling bonds or twisted bands. Fig. 2(g) and 2(h) exhibit the interfacial structure (dotted red ovals) between domain a and the substrate and the interface (dotted red rectangles) between domain m and the substrate, respectively, which clearly shows the atom registry from the substrate (Mg) to the film (Zn). The corresponding atomic modes of the interfacial structure are illustrated in Fig. 2(i). However, due to the mismatch between ZnO film and MgO substrate and the different surface atomic structure characteristics of the ZnO-(0001) plane and MgO-(001) plane, some dislocations and/or defects are expected at the interfaces. The high resolution TEM image in Fig. 2(b), for example, confirms the formation of the V-shape defects in the vicinity of the interface (indicated by the white dashed lines), probably originating from the difficulty of bonding between the polar plane and MgO (001) surface in the interface. Since the static barrier energy of each Mg atom in the (001) plane is only about 0.3 eV,[34] they could be easily detached from the surface, resulting in the formation of the defects when the substrate is heated during the growth. In reality, it has been reported [35] that point defects are indeed found in the MgO (001) surface. Another interesting aspect to examine is the boundary interface between the two domains in the ZnO film. As shown in Fig. 2(j), at the domain interface, there appears stacking faults only along the c-axis. In the direction perpendicular to the c-axis, the atoms seem to arrange smoothly across the domain interface. Figure 3 shows the cross-sectional TEM images from the 100℃ sample, also on the zone axis of MgO [110]. The thickness of the 100℃ sample is about 175nm, as shown in the low magnification TEM image (Fig. 3(a)), indicating that the ZnO growth at lower substrate temperature is faster than hat at higher substrate temperature. However, such a growth leads to polycrystalline ZnO film, as indicated by the SAED pattern in Fig. 3(b), which can also be seen from the high resolution TEM image in Fig. 3(c). Figures 3(d) and (e) are the corresponding amplified images of the rectangle regions marked by red lines in Fig. 3 (c). The high resolution TEM images reveal the evolution of the growth direction: Initially, the growth is along ZnO [0001] with double domains (a and m), as in the case of the growth at 250℃; when the thickness of the thin film reaches about 10 nm, as marked by the red dotted lines, the growth direction mutates into multiple growth directions, as labeled by dotted red circles in Fig. 3(c). Further analysis on the high resolution images in Fig. 3(d) and others (not shown) from the 100℃ sample shows that, the ZnO variants are mainly along the {0-332} group direction, slightly titled from the [10-11]* direction, whose corresponding XRD peaks appears in Fig. 1. It can also be seen from the TEM images that the portion of the thin film grown along near the [10-11]* direction is much larger than that along [0001] direction, in agreement with the previous XRD analysis. For the initial growth along [0001], the boundary structure of the double domains indicated by the blue lines in Figs. 3(c) and (e), parallel to the growth direction, is similar to the growth at 250℃. Furthermore, there appears additional boundary between domains a and m perpendicular to the growth direction, as labeled by the red arrow in Fig. 3(d).

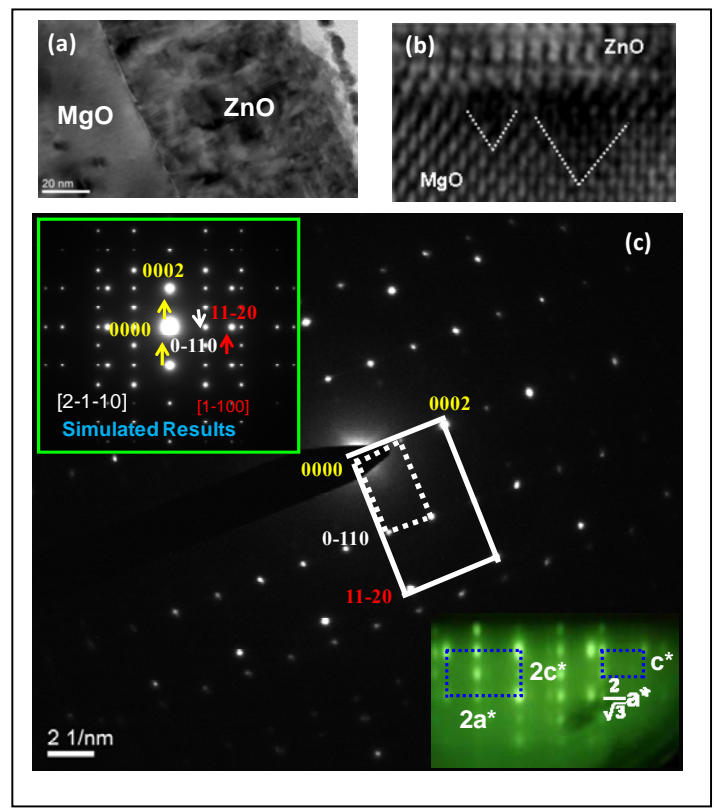

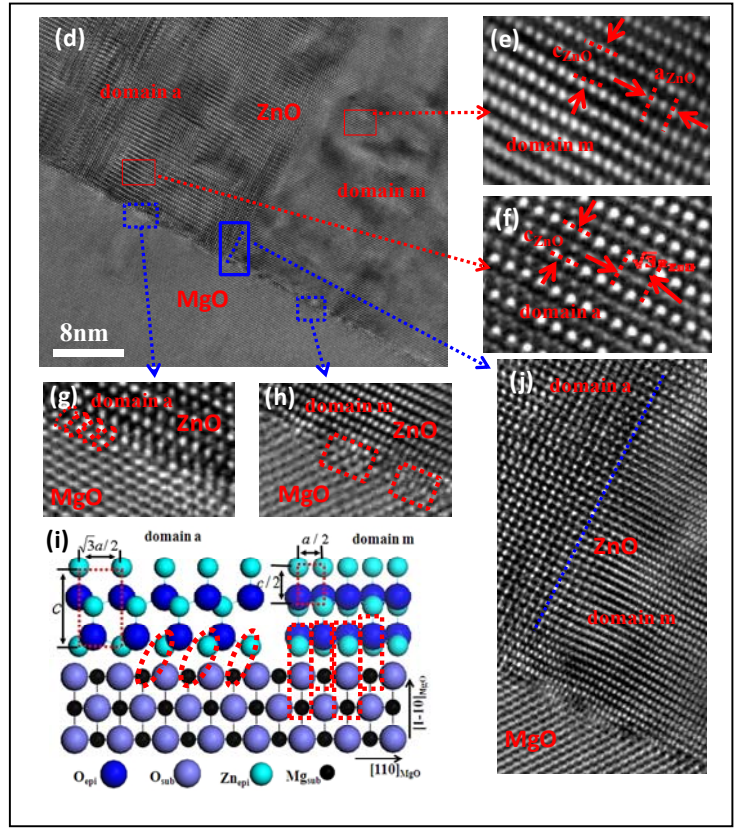

FIG. 2. (a) Low magnification TEM image of the ZnO/MgO interface from the 250℃ sample; (b) Further magnified image from the interface region, which shows "V-types" defects; (c) SAED patterns along the [110] MgO azimuth, showing diffraction spots from both the MgO substrate and the ZnO film with the double-domain film feature, consistent with the observation from the RHEED patterns and the calculated EDP image as shown in the lower-right inset and the upper-left inset; (d) Higher resolution TEM image of the ZnO/MgO interface from the 250℃ sample, showing two kinds of domains; (e) and (f) Further magnified images of the rectangle areas (marked by red solid lines) from the two domain areas in (d) showing the lattice parameters *a* and *c* marked by red dotted lines equal to 3.255 Å and 5.243 Å, respectively; (g) and (h) The amplified images of the rectangle areas labeled by blue dotted lines, showing the positional relationship between Zn atoms and Mg atoms in the two kinds of domains labeled by the dotted circle and rectangle lines, respectively; (i) The corresponding atomic model of the interfacial structure from the side-view along the [110] MgO azimuth; (j) The interfacial structure between the two kinds of domains, as labeled by the blue dotted line.

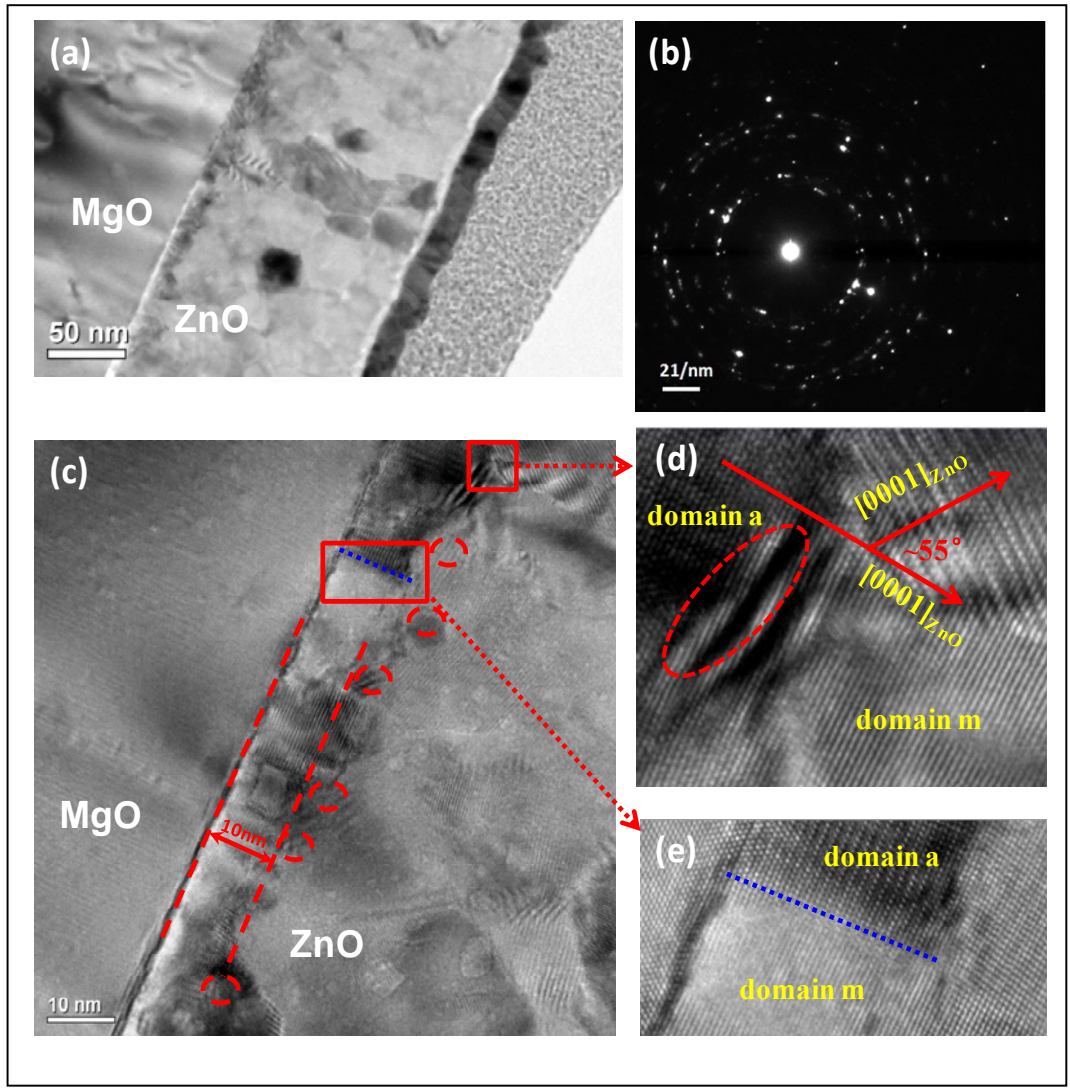

FIG. 3. (a) Low magnification TEM image of the ZnO/MgO interface from 100℃ sample; (b) The corresponding SAED pattern showing polycrystalline feature of the ZnO films; (c) High resolution TEM image of the ZnO/MgO interface from the 100℃ sample showing that the initial growth along [0001]azimuth is of about 10 nm, as labeled by red dotted lines, with the later multiple growth directions labeled by red dotted circle lines; (d) and (e) The amplified images of the rectangle areas from the ZnO film, which show the mutation of growth direction of the ZnO film from [0001] azimuth to [10-11] azimuth labeled by red arrows and its nanorod-like columnar growth labeled by red dotted line, respectively; (f) and (g) STEM images from the 250°C sample and 100℃ sample, respectively, indicating that the ZnO film grown at 250°C does not shown growth mutation, while the film grown at 100℃ exhibits mutation when the thickness is larger than about 10nm.

It is an interesting observation that, at the initial growth stage, the growth at 100°C and 250°C follows the same growth direction towards [0001] with the same double-domain feature. This is further confirmed through STEM images as described by Figs. 4(a) and (b), which correspond to the 250°C and 100°C samples, respectively. The results indicate that, at the initial stage, the substrate structure plays the most important role in the determination of film structure, which follows the rotational domain rule in heteroepitaxy predicted by Grundmann *et al*,[32] regardless of the growth temperature. Beyond the film thickness of about 10 nm, the growth at 250°C remains the same fashion as in the initial stage, while the growth at 100°C turns to other directions, as illustrated by the images of TEM in Fig. 3(c) and STEM (indicated by the circles) in Fig. 4(b). This could be due to the slower diffusion rate at lower temperature, which leads to localized growth orientation influenced by stress. And the reason why the mutation chooses to undergo towards the [0-332] variants could be possibly attributed to the angle of incidence flux. In fact, previous reports demonstrated that the growth direction of films can be controlled by the incidence beam flux.[36-38] Furthermore, through analyzing other high resolution TEM images (not shown here), the values of lattice parameters ($a$ = 3.267Å, $c$ =5.234Å), along [0001] were found different from those retracted along [0-332] directions ($a$ = 3.392Å, $c$ = 5.571Å) in the same thin film grown at 100°C. The $a$ and $c$ values determined from the growth areas along [0001] are 0.6% and 0.5% larger than the ideal bulk values ($a$ = 3.249Å and $c$ = 5.207Å), respectively, and those determined from the growth areas along [0-332] are 4.4% and 7.0% larger than the ideal bulk values, respectively. This is consistent with the XRD result that the 10-11 peak position is 36.15°, smaller than the theoretical value (36.25°).

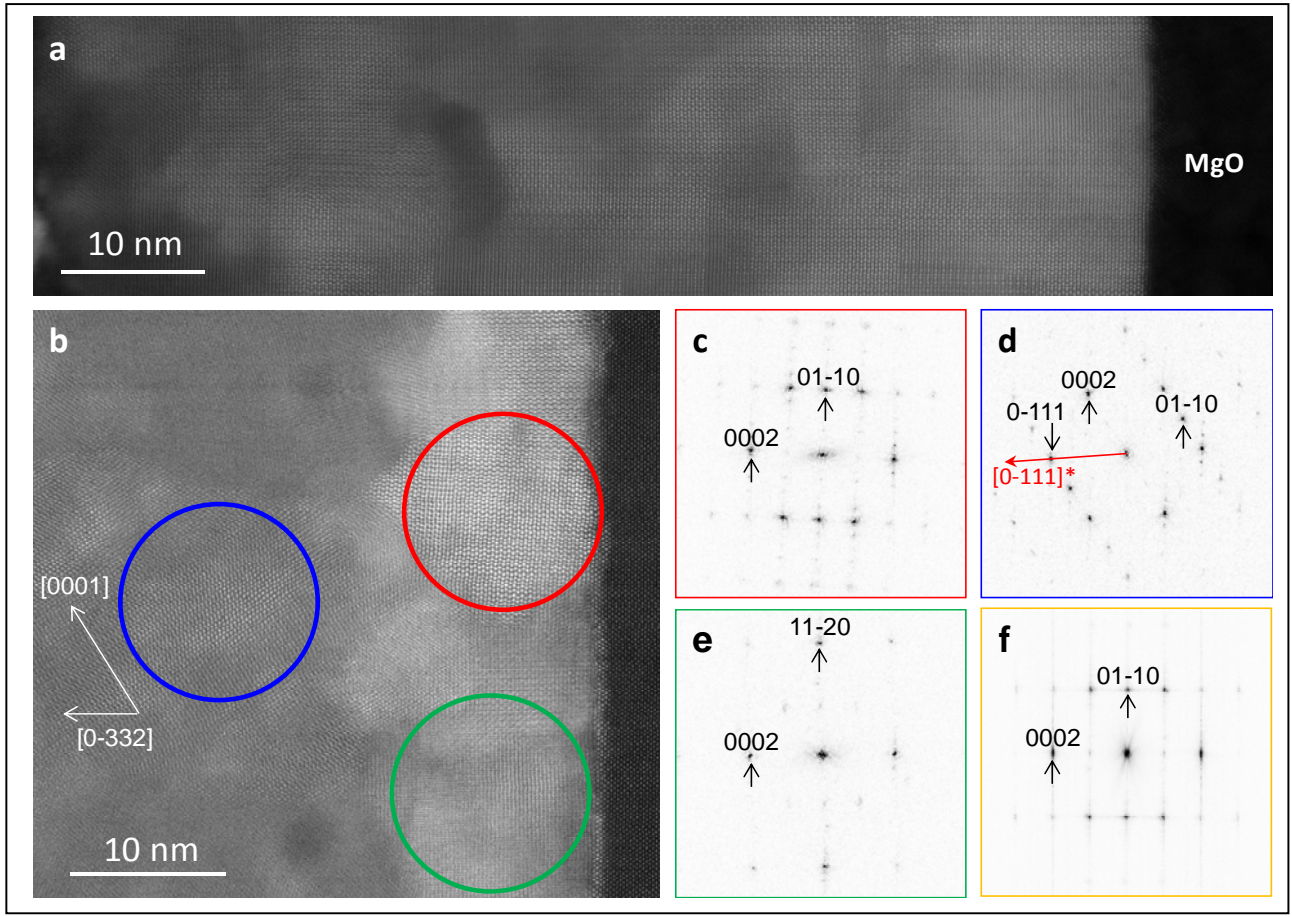

FIG.4. STEM images of the ZnO films grown at 250°C (a) and 100°C (b) respectively, showing the same interfacial structure between the films and the substrate, with different growth orientations beyond the interface region, as illustrated by the yellow dashed circle in (b) image. (c)-(e) Diffractograms from (c) red, (d) blue, (e) green circled area in (b). The red arrow in (d) points to [0-111]* direction which is the normal of (0-111) plane and deviates azimuth about 3°. (f) is the diffractogram from the film area in (a) for comparison.

### B. Optical and electronic properties of the ZnO thin films

In order to investigate the effect of substrate temperature on the optical properties of the ZnO thin films, their transmissivities are measured as shown in Fig. 5, with the transmissivity quantity of the MgO substrate set as 100%. Obviously, the transmissivity of the visible light is reasonably high for ZnO films grown at higher substrate temperature of 300°C, 250°C and 200 °C and tends to decrease for ZnO films grown at lower substrate temperature of 150°C and 100°C as shown by dark dashed circles.

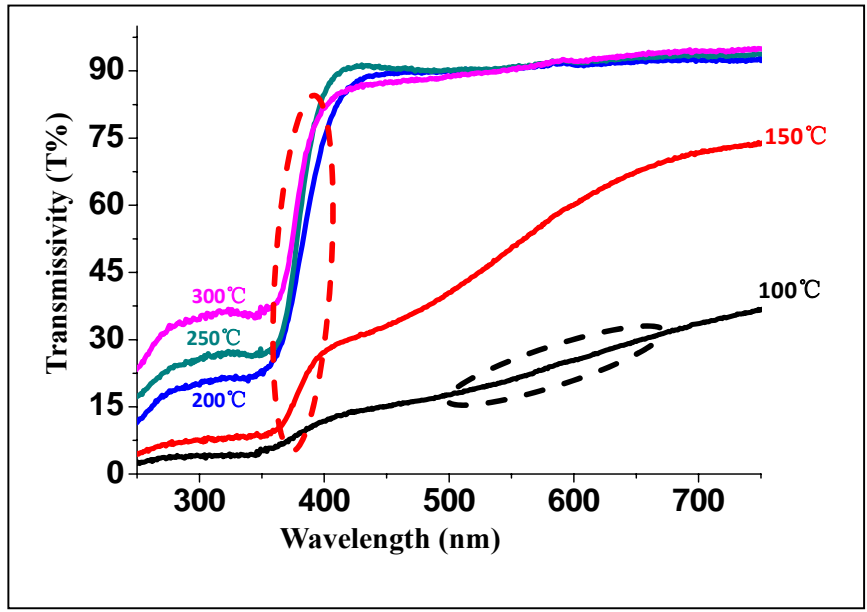

FIG.5. The transmissivity of the films. The black dotted circle for the film grown at 100 °C indicates that its absorption of nearly the entire visible light region. The red dotted circle corresponds to the band-gap feature of the films.

The evolution of the band-gap structure is also indicated by the red dashed circles. The possible reason leading to these results is that, the films grown at higher substrate temperature result in better crystallization, leading to bulk-like energy band-gap structure of ZnO; whereas the films grown at lower substrate temperature exhibit large disorders between grain boundaries, leading to different energy band-gap structure from bulk ZnO, such as a wider conduction band as reported before [39] as well as wider band tail, and the occurrence of a great number of defects also increases the absorption of visible light.

The change of band-gap structure is related to bond length modulations, which can be probed by extended X-ray absorption fine structure (EXAFS) spectra from the synchrotron-based equipment. The Zn-edge EXAFS results for ZnO films grown at 100 °C, 150 °C, 200 °C and 300 °C are presented at Fig. 6, which shows the electronic structure evolution of the ZnO films from multiple growth directions (for growth at 100 °C and 150 °C) to single growth direction with two rotational domains (for growth at 200

°C and 300 °C). The EXAFS spectra are measured using fluorescence mode, which utilizes a 32-elements Ge solid state detector with a digital X-ray processor system. Incident x-ray energy was selected with a three-quarters tuned Si(111) double monochromator, and the incident angle of X-ray beam was 45° as illustrated in the upper-left inset of Fig. 6(a). Although the grain sizes of the films are different (Table I), which induces a tremendous difference for the interfacial structure grown at low and high growth temperatures, the absorption peaks of all these film samples are well aligned, except that the intensity of absorption peaks from the films grown at 100°C and 150°C is a little smaller than grown at 200°C and 300°C. The results reveal no distinction of electronic structure at the bottom of the conduction band for the ZnO thin films even when the grain size becomes very small. Previously, Deok-Yong Cho et al [40, 41] reported that the structural disorders in amorphous ZnO thin films could induce the localization of the conduction band through the limited hopping interactions and orbital hybridization between adjacent ions as shown x-ray absorption spectra. Such phenomena do not appear in Fig. 6(a), indicating that the localization of the conduction band may not result from the interfacial effect of grain boundaries. In fact, Oba *et. al.* [42] demonstrated that the disorder bond and dangling bond between grain boundaries did not necessarily induce electronic states in the band-gap due to the domination of the Zn *4s* character at the bottom of the conduction band using an *ab initio* plane-wave pseudo-potential method. In Fig. 6(a), the amplified profiles of X-ray absorption near edge structure (XANES), as shown in the upper-right inset, exhibit mostly the $4p_{xy}$ empty states, because of the large ratio of $4p_{xy}$ states to $4p_z$ states. Besides, in the ZnO films grown at higher temperatures of 200 °C and 300 °C, there appear slightly stronger $4p_z$ states than grown at lower temperatures of 100 °C and 150 °C. This could be due to the angle change between the incident X-ray beam and the crystalline orientation. For the growth along [0001] at 200 °C and 300 °C, the angle between the incident X-ray beam and c-axis is about 45°; for the growth along the [0-332] direction, that direction changes to about 80°, which only presents $4p_{xy}$ empty states. The lower-right inset of Fig. 6(a), showing the calculated results of density of states (DOS) and theoretical XANES of ZnO using first principles full-potential

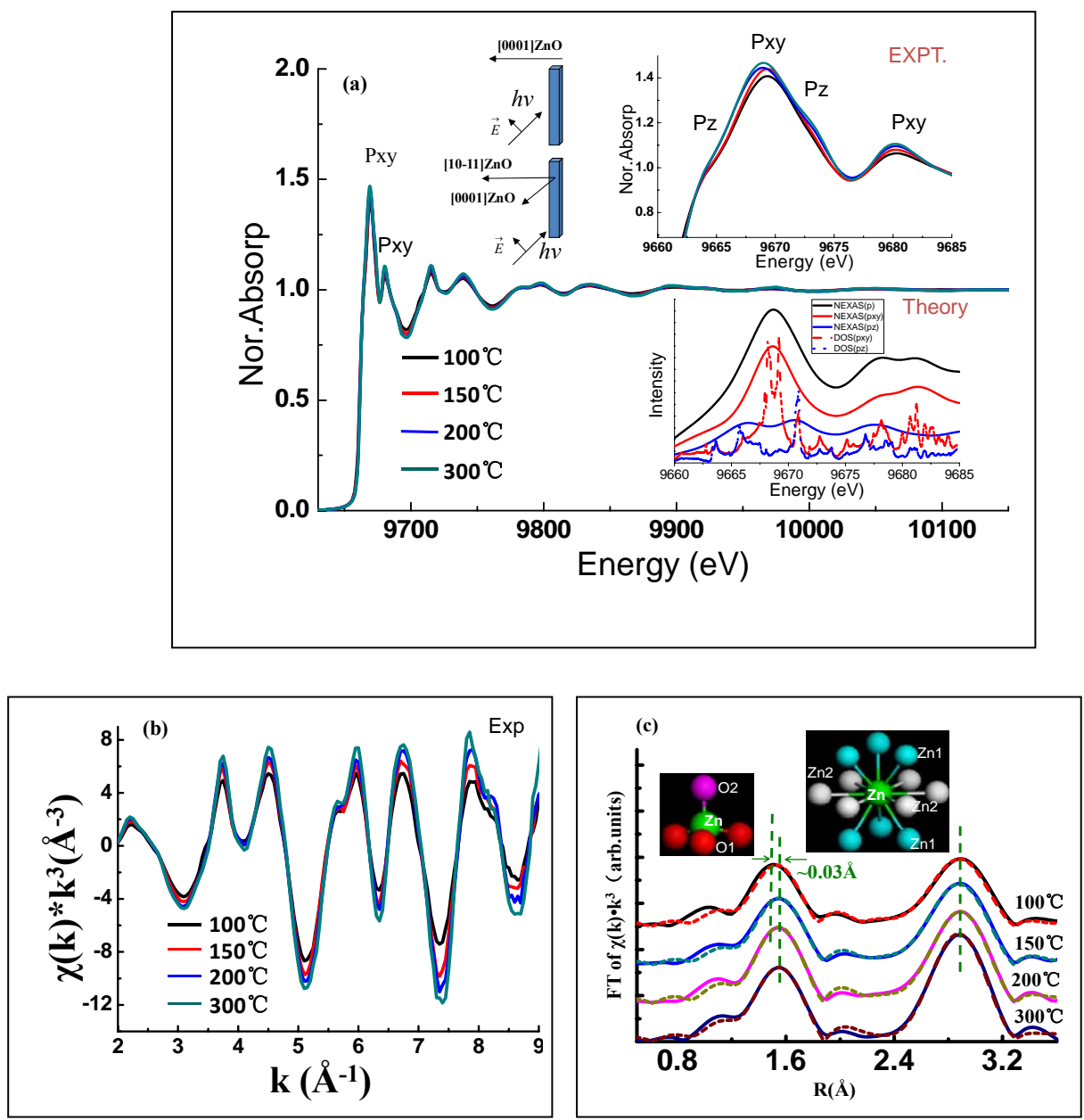

FIG. 6. Zn K-edge X-ray absorption spectra of four ZnO films. (a) The measured full X-ray absorption spectra. The upper-left inset shows the direction of incident x-rays and the possible orientations of crystalline grains. The upper-right inset corresponds to the zoom-in NEXAS spectra within a narrower energy range (from 9660 eV to 9685 eV), with indication of Zn electronic orbital components, namely, $p_{xy}$ and $p_z$. The NEXAS spectra for Zn K edge, including its $p_{xy}$ and $p_z$ components (i.e., NEXAS ($p_{xy}$) and NEXAS ($p_z$)), as well as $p_{xy}$ and $p_z$ projected density of states (i.e., DOS ($p_{xy}$) and DOS ($p_z$)), predicted from theory are also shown in the lower-right inset. (b) $\chi(k)*k^3$ as a function of $k$. The gradually enhanced amplitude reveals the gradually enlarged coordination number of Zn atom for growth from 100 °C to 300 °C. (c) Magnitude of Fourier transformed EXAFS using a Hanning window with a windowsill width of 0.5 Å. Data in the range of r=1-3.8 Å were used for the fitting. The insets show the corresponding atomic geometry structures of Zn-O1, Zn-O2, Zn-Zn1 and Zn-Zn2 bonds.

calculations based on density functional theory (DFT) [26], reveals the $4p_z$ states at the peak positions of 9666 eV and 9671 eV, demonstrating the weaker $4p_z$ features for ZnO films grown at 100 °C and 150 °C, compared to those grown at 200 °C and 300 °C. This is due to the twisting of growth directions from [0001] to [0-332] in some regions of films grown at 100 °C and 150 °C. In fact, the absorption spectra indicate some partial growth direction with a tilted angle about 45° along the normal of the substrate, consistent with above discussions. Figure 6(b) shows the $k^3$-weighted $\chi(k)$ signal of EXAFS data, the intensity of which increases with substrate temperature, revealing that the coordination number of Zn atoms increases with substrate temperature. The EXAFS data were Fourier transformed to $r$ space and fit to the theoretical EXAFS calculations using the IFEFFIT package, [27-29] as shown in Fig. 6(c). To minimize uncertainty, only the EXAFS data in the $k$ range of 2.0-12.427 Å$^{-1}$ were used for further analysis. The fitting included single- and multi-scattering paths and 95% polarization of the incident X-ray was taken into account in the data analysis. [43, 44] The dotted lines in Fig. 6(c) show the magnitudes of the Fourier transformed EXAFS data from the ZnO films. Note that the peaks are shifted by about 0.5 Å on the $r$ axis from their actual bond lengths due to the phase shift of the backscattered photoelectrons. The data were fitted with a fully occupied model of wurtzite structure, varying the parameters of bond length $R$, coordination number $N$ and Debye-Waller factor $\sigma^2$ (which includes thermal vibration and static disorders and is set the same for Zn-O1 and Zn-O2 and the same for Zn-Zn1 and Zn-Zn2). The fitting results are summarized in Table II. Combined with the results in Table I, it can be seen that the value of $\sigma^2$ becomes lower as the grain size increases, while the

TABLE II. Coordination number (N), bond length R and Debye-Waller factor ($\sigma^2$) of ZnO thin films grown at different substrate temperatures, determined through the fitting of orientation-dependent EXAFS data measured at the Zn K edge. $S_0^2$ is fixed at 0.95 [Refs. 38, 39] for the all the data fittings. For the model calculation, a fully occupied wurtzite structure (space group: p63m) with a=3.2495 Å, b= 5.2069 Å was used.

**Zn-O1**

| Parameters | Model | 100 °C | 150 °C | 200 °C | 300 °C |
|---|---|---|---|---|---|
| N | 3.0 | 2.1(±0.2) | 2.3(±0.2) | 2.4(±0.2) | 2.5(±0.2) |
| R(Å) | 1.97 | 1.94(±0.01) | 1.95(±0.01) | 1.97(±0.01) | 1.96(±0.01) |
| $\sigma^2$(Å$^2$) | | 0.0043(±0.0016) | 0.0038(±0.0012) | 0.0031(±0.0012) | 0.0036(±0.0014) |

**Zn-O2**

| Parameters | Model | 100 °C | 150 °C | 200 °C | 300 °C |
|---|---|---|---|---|---|
| N | 1.0 | 0.7(±0.1) | 0.8(±0.1) | 0.8(±0.1) | 0.8(±0.1) |
| R(Å) | 1.99 | 2.10(±0.03) | 2.08(±0.03) | 1.98(±0.03) | 1.98(±0.01) |
| $\sigma^2$(Å$^2$) | | 0.0043(±0.0016) | 0.0038(±0.0012) | 0.0031(±0.0012) | 0.0036(±0.0014) |

**Zn-Zn1**

| Parameters | Model | 100 °C | 150 °C | 200 °C | 300 °C |
|---|---|---|---|---|---|
| N | 6.0 | 5.1(±0.6) | 5.4(±0.5) | 5.8(±0.4) | 5.9(±0.5) |
| R(Å) | 3.21 | 3.18(±0.01) | 3.17(±0.01) | 3.19(±0.01) | 3.20(±0.02) |
| $\sigma^2$(Å$^2$) | | 0.0130(±0.0013) | 0.0113(±0.0009) | 0.0105(±0.0008) | 0.0094(±0.0008) |

**Zn-Zn2**

| Parameters | Model | 100 °C | 150 °C | 200 °C | 300 °C |
|---|---|---|---|---|---|
| N | 6.0 | 5.1(±0.6) | 5.4(±0.5) | 5.8(±0.4) | 5.9(±0.5) |
| R(Å) | 3.25 | 3.27(±0.01) | 3.28(±0.01) | 3.23(±0.01) | 3.24(±0.02) |
| $\sigma^2$(Å$^2$) | | 0.0130(±0.0013) | 0.0113(±0.0009) | 0.0105(±0.0008) | 0.0094(±0.0008) |

coordination number of the Zn ions increases gradually for bigger grains sizes, which is consistent with the $K^3 \cdot \chi(k)$ EXAFS data. The explanation is again related to the limited diffusion rate of adatoms for the substrate temperature below 150 °C, which hinders the expansion of the islands, resulting in lots of dangling bonds and thus increases the disorders of thin films. Especially, the film grown at 100 °C shows a large degree of disorder with a much smaller coordination number of Zn atom, close to those of amorphous ZnO films. In addition, the huge amounts of interfaces between grains, resulting from the mutation of the growth orientation and the tilted epitaxy, as indicated by STEM images in Fig. 4(b), could also lead to the increment of the disorders and the reduced coordination number of the Zn atoms. According to the equivalency of atomic geometry (as shown in the insets of Fig. 6(c)), two sets of Zn-O bonding (i.e. Zn-O1 and Zn-O2) and two sets of Zn-Zn bonding (i.e. Zn-Zn1 and Zn-Zn2) were considered for fitting. There is no obvious change of the values of R(Zn-Zn1) for ZnO films grown at the temperature range from 100 °C to 300 °C, while the values of R(Zn-O2) and R(Zn-Zn2) from the 100 ℃ and 150 ℃ samples are slightly larger than those grown at 200 ℃ and 300 ℃ and the corresponding values of R(Zn-O1) shows a contrast trend. The fitting results are consistent with the our observations from the high resolution TEM images that the structural parameters *a* and *c* from the 100 ℃ sample grown along [0-332] direction in some localized regions are 3.7% and 7.0% larger than the corresponding theoretical values, respectively. Besides, because the statistical possibilities of Zn-O1 bond is about three times that of Zn-O2 bond, when the Zn-O1 bonds become shorter and the Zn-O2 bonds become longer, the overall results would be that the nearest Zn-O bond length becomes shorter, leading to the left-shift of the peak position as indicated by green dashed lines in Fig. 6(c). These results also imply that the σ-bonds (also called bilayer bonds) [45-46] become slightly shorter and yet the π-bonds (also called c-axis bonds) [45-46] of ZnO become slightly longer in the region grown along [0-332] azimuth at lower growth temperatures. This modification of σ-bond or π-bond is not resolved from XRD data, probably due to the limited resolution of XRD, which detects the average structural information. As the $p_{xy}$ and $p_z$ states are very sensitive to bond length, the extended or shortened bonds will result in the transfer of electronic states. That is probably another reason why we observed the narrower width of the near Zn-k edge absorption of ZnO films grown at 100°C and 150°C, compared to those grown at 200°C and 300°C, as illustrated in the lower-right inset of Fig. 6(a). Furthermore, these can also contribute to the change of band structure, which is sensitive to bond length modulation. As a result, the transmissivity of the film grown at 100 °C is much smaller than grown at 200 °C and 300 °C, as shown in Fig. 5. It is thus concluded from the above discussions that, tilted epitaxy of ZnO thin films is a disadvantage for film quality.

Our work not only shows that the electronic structure of ZnO thin film depends on the grain size and growth direction, but also indicates that the control of growth conditions can tailor the grain sizes and the growth direction of the ZnO thin films, and thus tailor the electronic structure of the ZnO thin films. The investigation of grain-size dependence of ZnO nanostructures is of importance, as in the recent decades, ZnO nanorods and nanowires have shown particularly interesting properties and applications. For example, they can be used to tune electronic and optoelectronic devices that involve UV lasing action. [47,48] Knowledge of the electronic structure of nanorods or thin film grains is crucial to understand the basic physics for these applications.

## IV. CONCLUSIONS

In summary, the growth of wurtzite ZnO (0001) films on cubic MgO (001) substrates by molecular beam epitaxy is reported. For substrate temperature below 150 °C, the growth orientation of ZnO thin films exhibits mutation from single polar growth direction to multiple growth directions while for substrate temperature from 200 °C to 300 °C, the orientation is only along [0001], based on XRD and (S)TEM analysis. For the growth along [0001], there always appear two preferred in-plane rotational domains with a twist angle of 30º, and the interface relationship follows as: (0001)ZnO ∥ (001)MgO；

(01-10)ZnO ∥ (-110)MgO；[2-1-10]ZnO ∥ [110]MgO and (0001)ZnO ∥ (001)MgO； (11-20)ZnO ∥ (1-10)MgO； [1-100]ZnO ∥ [110]MgO. Besides, V-type defects are identified in the interface area resulting from the small barrier energy for Mg atoms in the (001) plane. For the mutation orientation of the films, due to the large ratio of surface to bulk, there appear higher disorder degree and smaller coordination number as compared to the growth orientation only along [0001]. Besides, for the region of thin film grown along [0-332] azimuth at the lower growth temperature, the lattice parameter of $a$ and $c$ are determined to be 3.392Å and 5.571Å, respectively, which are 4.4% and 7.0% larger than theoretical value, respectively. These phenomena also lead to the dramatic change of band-gap structure, as illustrated by the transmissivity measurements. However, the X-ray absorption spectra reveal that the large disorder of the ZnO films between grain boundaries does not necessarily lead to the distinctive electronic states at the bottom of the conduction band as observed in amorphous ZnO films.

## ACKNOWLEDGEMENTS

This work is supported by Natural Science Foundation of China (Grant Nos. 11204253, U1232110 and U1332105), the Specialized Research Fund for the Doctoral Program of Higher Education (Grant No. 20120121110021), the Fundamental Research Funds for Central Universities (Grant Nos. 2012121012, 2013SH001), and the National High-tech R&D Program of China (863 Program, No. 2013AA050901). Research carried out (in part) at the Center for Functional Nanomaterials, Brookhaven National Laboratory, is supported by the U.S. Department of Energy, Office of Basic Energy Sciences, under Contract No. DE-AC02-98CH10886.